# RELIABILITY OF MOBILE AGENTS FOR RELIABLE SERVICE DISCOVERY PROTOCOL IN MANET

[1]Roshni Neogy    [2]Chandreyee Chowdhury    [2]Sarmistha Neogy*

[1] Dept. of Information Technology, Jadavpur University
[2]Dept. of Computer Science & Engineering, Jadavpur University
*sarmisthaneogy@gmail.com

*ABSTRACT*

*Recently mobile agents are used to discover services in mobile ad-hoc network (MANET) where agents travel through the network, collecting and sometimes spreading the dynamically changing service information. But it is important to investigate how reliable the agents are for this application as the dependability issues(reliability and availability) of MANET are highly affected by its dynamic nature.The complexity of underlying MANET makes it hard to obtain the route reliability of the mobile agent systems (MAS); instead we estimate it using Monte Carlo simulation. Thus an algorithm for estimating the task route reliability of MAS (deployed for discovering services) is proposed, that takes into account the effect of node mobility in MANET. That mobility pattern of the nodes affects the MAS performance is also shown by considering different mobility models. Multipath propagation effect of radio signal is considered to decide link existence. Transient link errors are also considered. Finally we propose a metric to calculate the reliability of service discovery protocol and see how MAS performance affects the protocol reliability. The experimental results show the robustness of the proposed algorithm. Here the optimum value of network bandwidth (needed to support the agents) is calculated for our application. However the reliability of MAS is highly dependent on link failure probability.*

## 1. INTRODUCTION

A mobile agent is a combination of software program and data which migrates from a site to another site to perform tasks assigned by a user according to a static or dynamic route [1]. It can be viewed as a distributed abstraction layer that provides the concepts and mechanisms for mobility and communication [2]. An agent consists of three components: the program which implements it, the execution state of the program and the data. An agent may migrate in two ways, namely, weak migration and strong migration [3]. The platform is the environment of execution. The platform makes it possible to create mobile agents; it offers the necessary elements required by them to perform their tasks such as execution, migration towards other platforms and so on.

Typical benefits of using mobile agents include [4]
- Bandwidth conservation: sending a complex query to the database server for processing.
- Reduced latency: a lightweight server can move closer to its clients
- Load balancing: loads may move from one machine to the other within a network etc.

The route of the mobile agent can be decided by its owner or it can decide its next hop destination on the fly. Here, we assume the underlying network to be a Mobile Ad Hoc Network (MANET) that typically undergoes constant topology changes, which disrupt the flow of information over the existing paths. Mobile agents are nowadays used in MANETs for various purposes like service discovery [5], network discovery, automatic network reconfiguration etc. But before mobile agent based applications become commercially available for MANET, reliability estimation of them is very essential. Because of motion and location independence [1], this environment itself introduces new aspects to reliability (in terms of continuity of correct service).

In [1] [6] we tried to address this issue. In [7] reliability estimation of service discovery [5] agents is considered along with few modifications to the basic mechanism [5] in order to reflect the dynamism of the underlying environment. But node mobility is not considered explicitly, the link connectivity is assumed to follow Non Homogenous Poisson (NHPP) distribution. Thus we did not consider parameters like pattern of node mobility, transient failure of the links etc.

In this paper node mobility is represented by mobility models. Three kinds of mobility are considered. Random movement of nodes is proposed in Random Waypoint Mobility Model [8](RWMM), node movement based on temporal dependency in Smooth Random Mobility Model [9] (SRMM) and node movement based on spatial dependency in Reference Point Group Mobility Model [10] (RPGM). Also multipath propagation of radio signals is considered according to two ray propagation model [11] as it is simple and widely used in literature. In reality, even when an agent finds a node to be connected to where it is now residing, an attempt to migrate to that node (host site) may fail because of transient link errors like frequency selective fading, heavy rainfall etc. This effect can be modelled as a Poisson process. Reliability estimation algorithm of [7] is modified to incorporate these changes. Also a metric is proposed to measure the performance of the service discovery protocol. This enables us to thoroughly study the effect of MAS reliability on service discovery. Our MAS reliability estimation model is not tightly coupled to the application (service discovery) but with minor modifications it can be applied to agents deployed by other MANET applications as well.

In the following section we discuss about the service discovery process using mobile agents in MANET. Then in section 3 state of art regarding this topic is mentioned. In section 4 our model is introduced that is designed to estimate reliability of the mobile agent based system (MAS). The next section (5) gives the experimental results followed by concluding remarks in section 6.

## **2. THE PROCESS OF SERVICE DISCOVERY**

A service can be regarded as any hard- or software resource, which can be used by other clients. Service discovery is the process of locating services in a network. The following methods are used to discover and maintain service data [12]:
- service providers flood the network with service advertisements;
- clients flood the network with discovery messages;

- nodes cache the service advertisements;
- nodes overhear in the network traffic and cache the interesting data.

The first one corresponds to passive discovery (push model) whereas the next one describes active discovery (pull model). The other two methods mentioned above are the consequences of the first two. While the push mechanism is quite expensive in terms of network bandwidth (in the context of MANET), the pull mechanism suffers from poor performance (longer response times). Moreover there are other factors to be taken into account such as the size of the network (no. of nodes), availability of a service (how frequently services appear and disappear in the network), and the rate of service requests. Traditionally static service brokers are used for sharing service information which is not suitable for MANET due to its inherent dynamic nature. So as in [5] mobile agents can be deployed for this purpose (looking for services offered) as the agents can migrate independently [13], behave intelligently [14] and can negotiate with other agents according to a well defined asynchronous protocol [15].

The service discovery protocol presented in [5] is taken to be the basis here. We first estimate the reliability of MAS where the agents are roaming around the underlying MANET, discovering various services provided by the nodes in MANET. To do this the algorithm [5] uses two types of agents – a static Stationery Agent (SA) and mobile Travel Agent (TA). The SAs are deployed on per node basis. On the contrary the TAs are deployed dynamically to collect and spread service information among the nodes in MANET. A TA prefers those nodes on its route which it has not yet visited but which are reachable via nodes it already knows. In order to enforce this TA Route algorithm is proposed in [5] that determines the next target migration site of a TA. The SAs are responsible for controlling the no. of TAs roaming around the network. Thus depending on the incoming agent frequency (that is, no. of agents TA visiting a node is said to be incoming agent frequency) of TA, an SA can either create or terminate a TA depending on network bandwidth. Larger the bandwidth more agents can be supported leading to better performance and probably improved reliability.

## 3. RELATED WORKS

Reliability analysis of MAS in ad-hoc network is a complicated problem for which little attention has been paid. Most of the work done in this area is related to distributed systems and distributed applications. But as pointed out in [16], features like scalability and reliability becomes critical in challenging environment with wireless networks. However the scalability/reliability issue of MAS has been highlighted in [17] although the work does not focus on resource constrained environments like MANET. Moreover this work does not take into account the specific task for which the agents are deployed. But this is very much important as route of a mobile agent primarily depends on the purpose for which it is deployed. However, we could not find any work that considers estimation of reliability of service discovery agents for MANET but we found the following.

### 3.1. Service Discovery in MANET

There are already some approaches for service discovery in MANETs. Some of them attempt to optimize flooding by reducing its overhead [18], but they stillcause a lot of traffic. In [12] some device and service discovery protocols are discussed along with the issues in MANET. However the work does not provide a detailed concrete solution to the problem of service discovery though it suggests possible use of mobile agents in discovering services. In [19] an overlay structure is used to distribute service information. Here the network is divided into groups of nodes and nodes share service information among the group members. Only if a service request is not addressed in the present group, then it will be forwarded to the adjacent group. But in highly dynamic scenario this group formation can become an overhead. To reduce such overhead in [5] mobile agents are used. But this work does not take into consideration the movement of nodes before an agent finishes its job. Moreover this algorithm expects the network to retain the same connectivity pattern while an agent is roaming around the MANET.

### 3.2. Reliability in MANET

Due to the analytical complexity and computational cost of developing a closed-form solution, simulation methods, specifically Monte Carlo (MC) simulation are often used to analyze network reliability. In [20], an approach based on MC method is used to solve network reliability problems. In this case graph evolution models are used to increase the accuracy of the resultant approximation.

But little has been addressed on the reliability estimation of MANETs. In [21] analytical and MC-based methods are presented to determine the two-terminal reliability for the adhoc scenario. Here the existence of links was considered in a probabilistic manner to account for the unique features of the MANET. However, there remains a gap in understanding the relationship between a probability and a specific mobility profile for a node. In [22] MC-based methods are presented to determine the two-terminal reliability for the adhoc scenario. This work is an extension of that in [21] by including directly, mobility models in order to allow mobility parameters, such as maximum velocity, to be varied and therefore analyzed directly. The methods in this paper will now allow for the determination of impacts of reliability under specific mobility considerations. As an example, one may consider the different reliability estimate when the same networking radios are used to create a network on two different types of vehicles. Here node mobility is simulated using Random Waypoint mobility model [8]. But this Random Waypoint model of mobility being a very simple one often results in unrealistic conclusions. Moreover none of this work focuses on mobile agents but only the 2-terminal [8] or all terminal network reliability.

### 3.3. Reliability of Mobile Agents

Little attention has been given to the reliability analysis of MAS. In [23], two algorithms have been proposed for estimating the task route reliability of MAS depending on the conditions of the underlying computer network. In [24], which is an extension of the previous work, a third algorithm based on random walk generation is proposed. It is used for developing a random static planning strategy for mobile agents. However, in both the works the agents are assumed to be independent and the planning strategy seemed to be static. So this work does not address the scenario where agents can change their routes dynamically. Moreover, it does not address the issue of node mobility in between agent migrations.

In [1] a preliminary work has been done on estimating reliability of independent mobile agents roaming around the nodes of a MANET. The protocol considers independent agents only. Node and link failure due to mobility or other factors is predicted according to NHPP. An agent is allowed to migrate to any node with equal probability. This may not be realistic as some nodes may provide richer information for a particular agent deployed by some application. In [6] the MAS is assumed to be consisting of a number of agent groups demanding a minimum link capacity. Thus, each agent group requires different channel capacity. Hence, different groups perceive different views of the network. In this scenario the reliability calculation shows that even with large no. of heterogeneous agent groups with differing demands of link capacity, the reliability of the MAS gradually reaches a steady state. Since the task (for example service discovery) given to an agent primarily controls its routes, it is an important aspect and must be considered while estimating reliability. But the nature of the task and hence the mobile agent's movement pattern is not considered in any of these works.

## 4. OUR MODEL

Though mobile agents (MA) are recently used in many applications of MANET including service discovery, dependability analysis of such applications is not much explored. In [7] reliability analysis of service discovery agents is attempted. Here the agents will tend to migrate towards the crowded portion of MANET to collect and fast spread service information. But the work does not consider many insights like the effect of node mobility or transient faults. Also performance of service discovery protocol with respect to agent's performance is not considered. In the present workthat is an extended version of [7], the effect of underlying environment on service discovery agents are considered. Our model is described in three parts - modelling of MANET, modelling of service discovery agents on MANET and reliability estimation of mobile agent system

### 4.1. Modelling MANET

We model the underlying network as an undirected graph G= (V,E) where V is the set of mobile nodes and E is the set of edges among them. Let the network consist of N nodes, thus |V|=N that may or may not be connected via bidirectional links. The following assumptions are made ([25][26]):

- The network graph has no parallel (or redundant) links or nodes.
- The network graph has bi-directional links.
- There are no self-loops or edges of the type $(v_j, v_j)$.
- The states of vertices and links are mutually statistically independent and can only take one of the two states: working or failed.

Modeling node mobility is also one of the big problems in MANET. Many mobility models are proposed that can address this issue based on a specific application scenario such as campus network or urban areas. RWMM [8] is widely used in this regard for its simplicity. Here a mobile node randomly chooses a destination point (waypoint) in the area and moves with constant speed on a straight line to this point. After waiting a certain pause time, it chooses a new destination and speed, moves with constant speed to this destination, and so on. Here we have chosen a linear velocity $v_i(t)$ and a direction $\varphi_i(t)$ and calculate the next destination point as

$$x_i(t+\Delta t) = x_i(t) + \Delta t * v_i(t) * \cos\varphi_i(t) \qquad (1)$$

$$y_i(t+\Delta t) = y_i(t) + \Delta t * v_i(t) * \sin\varphi_i(t) \qquad (2)$$

But Random Waypoint can result in a sharp turn or sudden stop when the differential rate of change of velocity is infinity which is not feasible in a practical scenario [9]. So to smoothen such change in velocity, and hence make the model more realistic, SRMM [9] model is also considered. It can be simulated as in [1]

$$x_i(t+\Delta t)=x_i(t)+\Delta t*v_i(t)*\cos\varphi_i(t)+0.5*a_i(t)*\cos\varphi_i(t)*\Delta t^2 \qquad (3)$$

$$y_i(t+\Delta t)=y_i(t)+\Delta t*v_i(t)*\sin\varphi_i(t)+0.5*a_i(t)*\sin\varphi_i(t)*\Delta t^2 \qquad (4)$$

In situations like battlefield or rescue work people work in groups and hence movement in a group is commonly observed in such situations. So we can also use RPGM[10] in our simulation. In that case the speed and direction (angle) of mobile nodes (MNs) would follow that of their leader, called a reference point. The velocity of the leader can follow RWMM again. If $v_{leader}$ and $\varphi_{leader}$ represent the speed and direction of movement of the reference point respectively then the speed ($v_i$) and direction ($\varphi_i$) of $MN_i$ can be calculated as follows [10]

$$v_i(t) := v_{leader}(t)+random()*SDR*v_{max} \qquad (5)$$

$$\varphi_i(t) = \varphi_{leader}(t) + random()*ADR*\varphi_{max} \qquad (6)$$

Here SDR and ADR are speed and angle deviation ratio respectively having the following relation 0<SDR, ADR<1. Hence position of $MN_i$ at (t+$\Delta$t) time instant can be estimated as

$$x_i(t+\Delta t) = x_i(t) + \Delta t*v_i(t)*\cos\varphi_i(t) \qquad (7)$$

$$y_i(t+\Delta t) = y_i(t) + \Delta t*v_i(t)*\sin\varphi_i(t) \qquad (8)$$

Thus the movement of the nodes are simulated in one of the above three ways. Now the received signal power is calculated according to two ray propagation [11] of radio signals to handle multipath propagation effect. It is supplemented in Shannon's theorem to calculate link capacity as in [6]. A failed link can be treated as a link having zero capacity. Link capacity also varies with time even if the adjoining nodes do not change their position thus featuring transient nature of the faults.

Thus an initial configuration would be assumed. Afterwards due to mobility few links may fail and still a few may be revived also according to NHPP considering the transient nature of the faults.

### 4.2. Modelling Service Discovery Agents on MANET

In this paper, we assume that our MAS (S) at a time instant has m(t) independent agents (Travel Agents in [5]) that may move in the underlying MANET. Here m(t) indicates the fact that the no. of TAs varies with time as an SA can kill TAs [5]. The reliability of (S) is defined as the probability that (S) is operational during a period of time [2]. Later we define reliability of an individual agent in this context. The commonly used terms are listed in Table 1.

In this scenario we can think of an agent as a token visiting one node to another in the network (if the nodes are connected) based on the strategy listed as TA Route Algorithm in [5]. But node mobility in between agents' journey was not considered in [5]. So we have made necessary modifications to make the service discovery process more suited to the dynamics of MANET. A TA starts its journey from an owner (where it is created by SA) and moves from one node to another according to the TA route Algorithm [5]. But this movement is successful if the two nodes are connected and there is no simultaneous transmission in the neighbourhood of the intended destination (taken care of by the MAC protocol). So, we associate a probability with the movement to indicate transient characteristics of the environment, since, for example, the routing table may not be updated properly or the link quality may have degraded so much (due to increased noise level) that the agents are unable to migrate. Thus, if an agent residing at node $MN_A$ decides to move to node $MN_B$ (connected to $MN_A$) then the agent successfully moves to $MN_B$ with probability $p_t$. Here $p_t$ denotes the problem of unpredictability mentioned above. For example, noise level may increase due to heavy rainfall. If at any time an agent finds all unvisited nodes to be unreachable, the agent waits and then retries. This step tolerates the transient faults (temporary link failure) as an agent retries after some delay and hence improves system performance. This is not considered in [5] but to make the service discovery process more suitable to the MANET dynamicity, transient fault tolerance becomes a necessity.

Table 1. Notation

| Terms | Descritption |
|---|---|
| $m(t)$ | No. of Agents in the system at time t |
| $\lambda_i(t)$ | task route reliability of $i^{th}$ agent in a step of simulation |
| $\lambda(t)$ | average reliability of all agents |
| $r_i(t)$ | probability that $m_i$ is working correctly at time t, that is, the individual software reliability of mi |
| $SP_{total}$ | total no. of service providers |
| $SP_{dis}^i$ | no. of service providers discovered by $i_{th}$ node |
| $SP_{collected}$ | No. of service providers discovered by an agent |
| $R_i(t)$ | instantaneous reliability of service discovery |
| $R_{service}^i$ | reliability of service discovery protocol till time T |
| Q | no. of simulation steps |

In this scenario we study the reliability of MAS (consisting of the TAs) with respect to the network status and its conditions (for example connectivity of the links, path loss probability etc.). Each agent is expected to visit all operating nodes in MANET in order to collect and spread service information. We have taken the failure probability (P) of the mobile nodes ($P_{Node}$) to be a variable of Weibull distribution [6].

Now reliability of MAS ($R_s$) can be defined as

$$R_s = \{R_{MAS}|R_{MANET}\} \qquad (9)$$

This is a conditional probability expression indicating the dependence of MAS reliability on conditions of MANET. Here reliability of MANET ($R_{MANET}$) can be treated as an accumulative factor of ($1-P_{Node}$) and $P_{Link}$. $P_{Link}$ can be treated as a combination of P ($p_r$ is at an acceptable level) and the mobility model. Here $p_r$ denotes the received power at node j after traversing distance $d_{ij}$ from sender node i. Here we calculate individual agent reliability on the underlying MANET as follows:

If an agent can successfully visit M nodes out of N(desired) then it has accomplished M/N portion of its task. Thus reliability in this case will be M/N.

But if the application requires all N nodes to be visited to complete the task and in all other cases the task will not be considered to be done, the calculation will be modified as:

If an agent can successfully visit all N nodes desired then it has accomplished its task. Thus reliability in this case will be 1. In all other cases it will be 0.

Above definitions of agent reliability works only if there is no software failure of the agent (assumed to follow Weibull distribution [6]).

Now, the probability that the MAS is operational i.e., reliability of MAS ($R_{MAS}$) can be calculated as the mean of reliability of all its components, that is, the agents in this system. Clearly it is function (m(t)) of time as the total no. of TAs present varies with time

$$R_{MAS}(t) = \frac{\sum \{Agent Reliabilities\}}{m(t)} \tag{10}$$

However the performance of the service discovery protocol is measured by a node's success in discovering service providers in the network. All or a few nodes in MANET may act as service providers, thus $SP_{total} <= N$. The instantaneous reliability of $MN_i$ ($R_i(t)$) can be given as

$$R_i(t) = \frac{SP_d^i}{SP_{to}} \tag{11}$$

If $R_i(t)$ is integrated over time, it gives the overall service discovery reliability $R_{service}^i$ as follows

$$R_{service}^i = \frac{1}{t} \int_0^t R_i(t) dt = \frac{1}{T} \sum_T R \tag{12}$$

Here time is divided in discrete time steps and snapshot of the system is taken in short periodic intervals. As this interval tends to 0 a near continuous picture of S can be observed. The same is repeated Q times according to Monte Carlo simulation [1]. The following algorithm presented in this paper describes the reliability analysis of agents deployed for discovery and spreading of service information among the nodes in MANET. Reliability of the agent system is calculated according to equation 10 while equation 11 along with12 calculates the performance of individual nodes.

### 4.3. Detailed steps of reliability estimation

*1) Input parameters*:   M (initial no. of TAs in the system), the initial state of the network (node position, location, speed of the nodes)
*2) Detailed Steps:*
    The TA determines its next target according to the following algorithm.
    *TA_Route_Mod()*
 1 First the travel agent tries to find yet unvisited nodes, which are common neighbors of previously visited nodes. These common neighbor nodes have highest priority because their services can be used directly by more than one node.
 2 If all such common neighbors have been visited then the agents will next visit the nodes not visited yet. If there are more than one unvisited neighbors, the mobile agent can choose to visit any one of them.

3  If there are no unvisited nodes in direct range of a mobile agent then a node with unvisited neighbors is revisited. If there are two such potential nodes, the node with the lowest RSN is chosen.

4  If the first three conditions fail, the node with the lowest RSN becomes the node visited next.

Node mobility is considered here and is detailed in the following algorithm. As such the agent's decision in choosing the next destination depends on the currently reachable set of nodes. Here we are hoping that at short time intervals the network topology will not show huge changes so as to render the common neighboring nodes absolutely disconnected or be reduced to a node with very few neighbors (less than the number of nodes which influence the MA's decision to grant it the highest priority).

*Reliability_Calculation()*

1. Initialize n (that is the no. of mobile nodes successfully visited by an agent) to 0 and a source for the mobile agent.
2. Input network configuration (V, E) in the form of an edge list.
3. Node mobility is simulated according to RWMM/SRMM/RPGM.
    3.1. Node connectivity received signal power ($P_r$) (according to Two ray propagation model) and hence node connectivity is calculated as in [14].
    3.2. Some nodes may also fail because of software/hardware failure or become disconnected from the network according to another NHPP (or uniform) distribution. Node failure can be simulated by deleting the edges e from E' further that are incident on the failed node.
4. According to Weibull distribution we find individual software reliability $r_i$ for an agent i.
5. Breadth First Search (BFS) is used unless all connected subgraphs are assigned a proper cluster id. Thus, an isolated node is also a cluster.
6. The agents perform their job on this modified graph according to *TA_Route_Mod()*.
7. Repeat step 6 for all agents (m(t)) in the system.
8. When an agent comes back its $SP_{collected}$ is used to find its reliability, $\lambda_i(t)$ according to equation 11 and 12.
    8.1. Spawn new agents if service information is still incomplete.
9. Repeat steps 3 to 7 until all nodes are visited or the new destination falls in a different cluster.

10. Calculate $$\lambda_i(t) = \frac{n}{N} \tag{13}$$

Here the value of n depends heavily on the conditions of the underlying network.

11. Reset the value of n.

12. Calculate $$\lambda(t) = \frac{1}{k}\sum_{i=1}^{k}\lambda_i(t)r_i \tag{14}$$

13. Repeat steps 3 to 12 Q (simulation steps) times.

14. Calculate node reliability $$\frac{1}{Q}\sum_{q=1}^{Q}\lambda(q,t) \tag{15}$$

It is to be noted that step 3 is repeated for every move of the agent to take care of network dynamicity. If an agent fails to move because of background noise level, then it may retry depending on the amount of delay that the respective application can tolerate.

## 5. RESULTS

The simulation is carried out in java and can run in any platform. The initial positions of MAs and the initial network configuration are read from a file. The default values of parameters are listed in Table 2. Unless otherwise stated, the parameters always take these default values. Detailed analysis of simulation results is shown.

In reality network dynamicity affects agent migration and hence the reliability of MAS is found to depend heavily on its size (no. of agents) particularly for bigger MANETs. This fact is shown in figure 1. This graph is taken for two different scenarios. In one of them, each node can receive maximum (maximum tolerated frequency) 20 agents (gray columns), additional agents would get killed resulting in a drop in agent reliability. For the other one, the maximum tolerated frequency is kept at 15. For m(t) (<=15) as far as the underlying MANET remains connected, all agents will be able to complete their job if there is no software failure in them as shown in figure. But if m(t) is increased any further, then agent reliability drops as it exceeds the maximum tolerated agent frequency. Higher the no. of agents tolerated in the network greater will be the overall agent reliability for bigger MAS. This justifies the right side of the graph in figure 1.

Table 2. Default values for simulation parameters

| Parameter Name | Value |
|---|---|
| Number of nodes(N) | 25 |
| Number of agents(m(t)) | 25 |
| Number of simulations(Q) | 200 |
| $SP_{total}$ | 2 |
| Maximum Agent Frequency tolerated(MAX FR) | 20 |
| Link Failure Probability | 0.2 |
| Time for each simulation run | 750min |
| Mobility Model | SRMM |

Every MANET has a bandwidth limitation that in turn restricts the maximum value of m(t) during a period. Thus, it can be observed that there is a maximum incoming agent frequency supported by a node (figure 2). Higher value of this indicates greater bandwidth provided by MANET. So as expected, with higher bandwidth ourMAS become more reliable. But it can be observed that with m(t)=25 when the incoming agent frequency reaches above 18, the MAS reaches an almost steady state with overall reliability of (around) 0.95. This gives the optimum value of bandwidth to be provided by MANET for this scenario.It is interesting to observe that for optimum performance all m(t) agents need not be tolerated at a time (as m(t)> maximum agent frequency tolerated).

Now if N is increased, the overall reliability reaches a steady state at N=20 as long as m(t) (<30) and hence is comparable to the agent frequency (=20) supported by the nodes (figure 3). Thus our approach is found to be scalable for MANETs as big as 45 nodes. But for large m(t) (>=30), nodes may kill appreciable number of agents resulting in a drop in reliability as a MANET grows. But this result indicates the scalability of the service discovery approach for crowded MANET as change in network size does not appreciably affect the reliability of MAS rather it helps the agents to discover service providers even faster by redundant routes.

The movement pattern of the nodes is found to play an important role in many MANET applications as it affects link existence probability. For smaller MANETs, nodes moving randomly according to RWMM exhibits better MAS reliability as compared to SRMM or RPGM (figure 4).

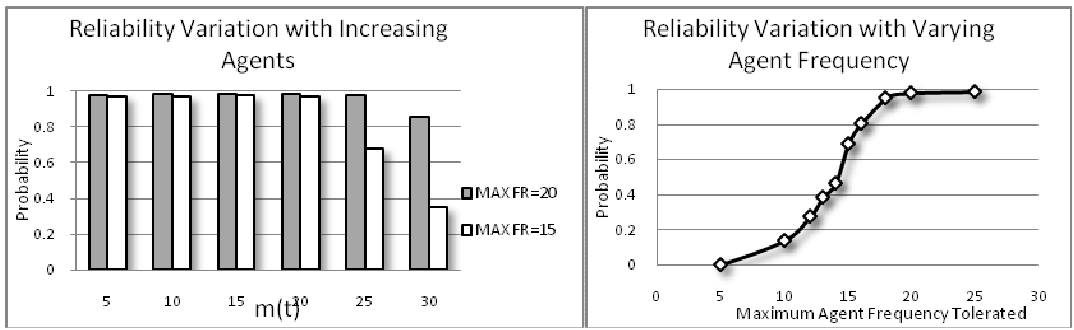

Figure1. Reliability variation with no.of agents      Figure2. Reliability variation with varying tolerance limit on accepted agent frequency

The reason can be periodicity in link existence probability [8]. But as a MANET grows, the performance of MAS becomes almost independent of the effect of different mobility models and hence user movement pattern. This result shows that the reliability prediction made in one scenario (say campus network following SRMM) is very much applicable in several other scenario (including disaster relief following RPGM).

Though mobility pattern of the nodes does not affect its agents but the transient failure of the links does. The link (transient) failure probability is found to affect reliability of the

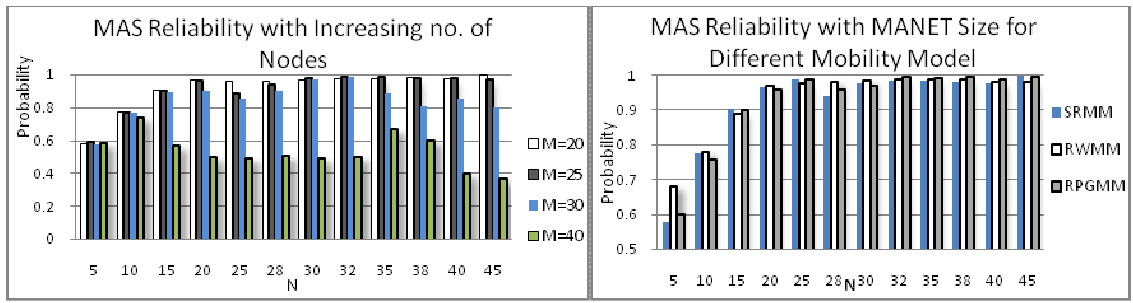

Figure 3.Reliability with varying no. of nodes(N)      Figure4. Reliability variation with MANET size for different mobility models

agents(figure 5). Thus even when the nodes are well within radio coverage of each other due to environmental effects like frequency selective fading or even heavy rainfall links may not be established. This is represented by the link failure probability (LFP), a metric that roughly

239

follows NHPP. As more links fail, the network becomes partitioned resulting in a sharp fall in MAS reliability (figure 5). It can be observed that when the link failure probability reaches above 0.3, the network graph loses connectivity and as a result some agents may never reach the nodes residing in another component of the network.

The reliability of service discovery protocol is analyzed according to equations11 and 12. The performance is measured from $MN_{24}$'s view. First the performance is measured with no. of service providers ($SP_{total}$). As $SP_{total}$ increases service availability increases and hence reliability of the protocol after initial perturbation reaches a steady state (figure 6). But the effect of service availability is more apt in MANETs with more transient disturbance. As the environmental conditions stabilize the agents perform more reliably discovering almost all service providers even when they are scarce.

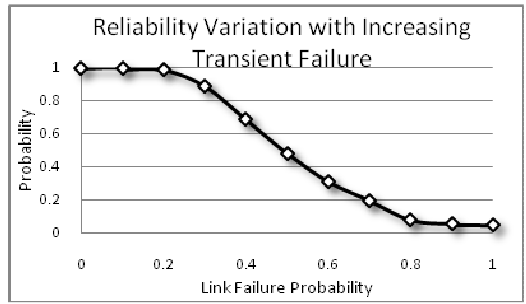

Figure 5.Reliability variation with varying link failure probability

The effect of MANET size on the performance of service discovery is also studied as shown in figure 7. But as a MANET becomes more crowded (N>30) redundant paths emerge to reach service providers even when they are scarce. Here the total no. of service providers is taken to be 13. The drop in reliability for LFP=0.5 near 25 to 30 nodes MANET is because at this stage the nodes newly added increases the MANET boundary rather than making it more crowded. Thus the overall network connectivity worsens. But with stable environment (low transient errors) the service discovery protocol manages to discover most service providers.

## 6. CONCLUSION

In this paper, a scalable approach to estimate the reliability of MAS for MANET is presented. The agents are deployed for collecting and spreading service information in the network. The reliability is found to depend heavily on MANET dynamics in terms of transient link failure and supported network bandwidth. However the effect of different movement patterns of the users is found to affect MAS reliability a little. Thus conclusion drawn for one scenario (represented by a mobility model) may well be applied to others (other mobility models) if the transient failure probability <0.3. Moreover our approach is found to be scalable for bigger MANETs too. The agents choose their destination on route according to a service discovery algorithm based on [5]. During the time an agent is visiting a node, the underlying network may change according to SRMM taken to be the default case.

The protocol is validated and results are shown in section 5. As can be seen, reliability improves heavily if the network supports higher no. of agents. Hence as per expectation, this modified model works well in an efficient manner regardless of the pattern of changes in the network topology.A metric is proposed to measure reliability of the protocol itself and shows how agents help in discovering services overcoming the effect of poor network connectivity.

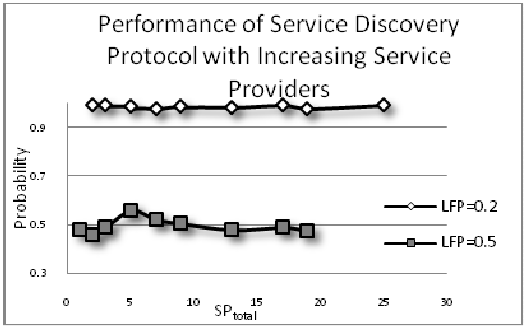 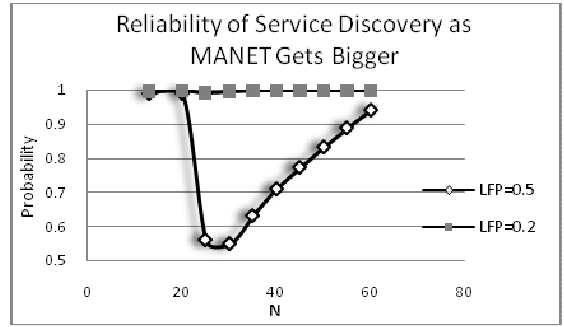

Figure 6.Performance of service discovery protocol with increasing service providers

Figure7. Performance of service discovery protocol with increasing N

The results of the simulation also corroborated our expectation. Our future work may include multiagent communication to make the process of service discovery even more reliable and efficient.We are also planning to include security issues in it.

## REFERENCE

[1] C. Chowdhury, S. Neogy, "Estimating Reliability of Mobile Agent System for Mobile Ad hoc Networks", Proc. 3rd International Conference on Dependability, pp. 45-50, 2010.

[2] J. Cao, X. Feng , J. Lu , S. K. Das, "Mailbox-Based Scheme for Designing Mobile Agent Communications", Computer, v.35 n.9, pp 54-60, September 2002.

[3] N. Migas, W.J. Buchanan, K. McArtney, "Migration of mobile agents in ad-hoc, Wireless Networks", Proc.11th IEEE International Conference and Workshop on the Engineering of Computer-Based Systems, pp 530 – 535, 2004.

[4] D. B. Lang and M. Oshima, "Seven good reasons for mobile agents", in the Magazine Communications of the ACM, Vol. 42, Issue 3, March, 1999.

[5] R.T. Meier, J. Dunkel, Y. Kakuda, T. Ohta, "Mobile agents for service discovery in ad hoc networks", Proc. 22nd International Conference on Advanced Information Networking and Applications, pp 114-121, 2008.

[6] C. Chowdhury, S. Neogy, "Reliability Estimate of Mobile Agent Based System for QoS MANET Applications", in the Annual Reliability and Availability Symposium 2011, pp. 1-6, 2011.

[7] C. Chowdhury, S. Neogy, "Reliability Estimation of Mobile Agents for Service Discovery in MANET", in the International Conference on Parallel, Distributed Computing technologies and Applications (PDCTA 2011), pp. 148-157, 2011.

[8] C. Bettstetter and C. Wagner, "The Spatial Node Distribution of the Random Waypoint Mobility Model", in the Proceedings of German Workshop on Mobile Ad Hoc Networks (WMAN) (2002).

[9] C. Bettstetter, "Smooth is better than sharp: a random mobility model for simulation of wireless networks", in the Proceedings of the Fourth ACM International Workshop on Modeling, Analysis and Simulation of Wireless and Mobile Systems, pp. 19-25, 2001.

[10] X. Hong, M. Gerla, G. Pei, and C. Chiang. "A group mobility model for ad hoc wireless networks". In Proceedings of the ACM International Workshop on Modeling and Simulation of Wireless and Mobile Systems (MSWiM), August 1999.

[11] M. Rooryck, "Modelling multiple path propagation- Application to a two ray model", in the journal of L'Onde Electrique , ISSN 0030-2430, vol. 63, pp. 30-34, Aug.-Sept. 1983.

[12] J. Albert, S. Chaumette, "Device and Service Discovery in Mobile Ad-hoc Networks", Technical report, Master 2 SDRP, Université Bordeaux 1, Jan. 16, 2007.

[13] M. Wooldridge, N. R. Jennings, "Intelligent agents - theory and practice", Knowledge Engineering Review 10 (2), pp. 115-152, 1995,.

[14] S. Ossowski, A. Omicini, "Coordination Knowledge Engineering", Knowledge Engineering Review 10 (2), pp. 115-152, 2002.

[15] J. Dunkel, R. Bruns: Software Architecture of Advisory Systems Using Agent and Semantic Web Technologies,Proceedings of the IEEE/ACM International Conference on Web Intelligence, Compiégne, France, IEEE Computer Society, pp. 418-421, 2005.

[16] O. Urra, S. Ilarri and E. Mena, "Agents jumping in the air:dream or reality", In 10th International Work-Conference on Artificial Neural Networks (IWANN'09), Special Session on Practical Applications of Agents and Multi-Agent Systems, pages 627–634. Springer, 2009.

[17] S. Ilarri, R. Trillo, E. Mena, "SPRINGS: A scalable platform for highly mobile agents in distributed computing environments", In: 4th International WoWMoM 2006 Workshop on Mobile Distributed Computing (MDC 2006), pp. 633–637. IEEE, Los Alamitos 2006.

[18] S. Helal, N. Desai, V. Verma, C. Lee: KONARK – A Service Dicovery and Delivery Protocol for Ad Hoc Networks. Proc. of the 3rd IEEE Conf. on Wireless Communication Networks (WCNC'03), Volume 3, pp. 2107-2113, 2003.

[19] M. Klein, B. Konig-Ries, and P. Obreiter. Lanes – A Lightweight Overlay for Service Discovery in Mobile Ad Hoc Networks. In Proc. of the 3rd IEEE Workshop on Applications and Services in Wireless Networks (ASWN2003), 2003.

[20] T.Elperin, I. Gertsbakh, M. Lomonosov, "Estimation of network reliability using graph evolution models", in the IEEE Transactions on Reliability, Vol. 40. No. 5. pp 572-581, 1991.

[21] J.L. Cook, J.E Ramirez-Marquez, "Two-terminal reliability analyses for a mobile ad-hoc wireless network", in Reliability Engineering and System Safety, Vol. 92, Issue 6, pp. 821-829, June 2007.

[22] J. L. Cook, J.E. Ramirez-Marquez, "Mobility and reliability modeling for a mobile ad-hoc network", IIE Transactions, 1545-8830, Vol. 41, Issue 1, pp. 23 – 31, 2009.

[23] M. Daoud, Q. H. Mahmoud, "Reliability estimation of mobile agent systems using the Monte Carlo approach", Proc. 19th IEEE AINA Workshop on Information Networking and Applications, pp 185–188, 2005.

[24] M Daoud, Q. H. Mahmoud, "Monte Carlo simulation-based algorithms for estimating the reliability of mobile agent-based systems" Journal of Network and Computer Applications, pp 19–31, 2008.

[25] ML. Shooman, "Reliability of computer systems and networks: fault tolerance, analysis, and design". New York: Wiley; 2002.

[26] C. Srivaree-ratana, A. Konak, A. E. Smith, "Estimation of all-terminal network reliability using an artificial neural network", Computers and Operations Research, Vol. 29, pp. 849–868, 2002.

**Authors**

**Roshni Neogy** is an undergraduate engineering final year student of Information Technology at Jadavpur University. Her research interests are in wireless and mobile networking, social network analysis.

**Sarmistha Neogy** is faculty in Jadavpur University at present and is in teaching profession since last eighteen years. She has been an active researcher in the areas of distributed systems, fault tolerance, mobile computing and security in wireless networks.

**Chandreyee Chowdhury** is a junior faculty in the department of Computer Science and Engineering at Jadavpur University. She received M. E in Computer Science and Engineering from Jadavpur University in 2005. Currently she is pursuing Ph. D under the guidance of Dr. Neogy. Her research interests include reliability and security in wireless networks.